\documentclass[conference]{IEEEtran}

\def\BibTeX{{\rm B\kern-.05em{\sc i\kern-.025em b}\kern-.08em
T\kern-.1667em\lower.7ex\hbox{E}\kern-.125emX}}

\usepackage[subtle,tracking=normal]{savetrees}

\IEEEoverridecommandlockouts
\usepackage{amsmath,amssymb}
\usepackage{graphicx}
\usepackage{color}
\usepackage{soul}
\usepackage{enumerate} 
\usepackage{subcaption}
\usepackage{cite}
\usepackage{multirow,tabularx}
\usepackage{cases}
\usepackage{ifthen}
\usepackage{wasysym}
\usepackage{stackrel}
\usepackage{bm}
\usepackage{algorithm}
\usepackage{algpseudocode}
\usepackage{flushend}
\usepackage{xcolor}
\definecolor{blackurlcolor}{rgb}{0,0,0}
\newcommand{\hide}[1]{\ifthenelse{\boolean{false}}{#1}{}}
\newcommand\norm[1]{\left\lVert#1\right\rVert}
\newcommand\abs[1]{\left\lvert#1\right\rvert}

\newcommand{\qed}{\nobreak \ifvmode \relax \else
      \ifdim\lastskip<1.5em \hskip-\lastskip
      \hskip1.5em plus0em minus0.5em \fi \nobreak
      \vrule height0.75em width0.5em depth0.25em\fi}

\newcommand{\barr}{\begin{array}}
\newcommand{\earr}{\end{array}}

\newcommand{\benum}{\begin{enumerate}}
\newcommand{\eenum}{\end{enumerate}}

\newcommand{\kroneck}{\otimes}

\newcommand{\bsp}{\begin{slide*}}
\newcommand{\esp}{\end{slide*}}
\newcommand{\bsl}{\begin{slide}}
\newcommand{\esl}{\end{slide}}

\makeatletter
\g@addto@macro\normalsize{\setlength\abovedisplayskip{3pt plus 1pt minus 1pt}\setlength\belowdisplayskip{3pt plus 1pt minus 1pt}\setlength\abovedisplayshortskip{2pt plus 1pt}\setlength\belowdisplayshortskip{2pt plus 1pt}}
\makeatother

\algrenewcommand\algorithmicrequire{\textbf{Input:}}
\algrenewcommand\algorithmicensure{\textbf{Output:}}

\IEEEoverridecommandlockouts

\makeatletter
\newcommand*\titleheader[1]{\gdef\@titleheader{#1}}
\AtBeginDocument{%
  \let\st@red@title\@title
  \def\@title{%
    \bgroup\normalfont\large\centering\@titleheader\par\egroup
    \vskip 0.5em\st@red@title}
}

\begin{document}

% Title
\title{HAPS-Complemented Terrestrial Networks}
\author{\IEEEauthorblockN{Animesh Yadav\IEEEauthorrefmark{1} and Halim Yanikomeroglu\IEEEauthorrefmark{2}}
\IEEEauthorblockA{\IEEEauthorrefmark{1}School of EECS, Ohio University, Athens, OH, USA, \IEEEauthorrefmark{2}Dept. of SCE, Carleton University, Ottawa, ON, Canada}
Email: yadava@ohio.edu\IEEEauthorrefmark{1}, halim@sce.carleton.ca\IEEEauthorrefmark{2}}

\maketitle

\begin{abstract}
% Add your abstract here
We consider a downlink multicell multiple-input multiple-output (MIMO) system in an urban region, with a focus on improving the capacity of cell-edge user equipments (UEs). These UEs typically experience lower rates than near UEs because of shadowing, path loss, and inter-cell interference (ICI). To address this issue, we integrate a high-altitude platform station (HAPS) with the terrestrial network as a relay for edge-UE transmissions. We assume that the HAPS operates in full-duplex (FD) mode and exploits its large physical size to enhance passive self-interference (SI) suppression by separating its transmit and receive antennas. In the proposed scheme, each terrestrial base station (BS) forwards edge-UE data to the FD-HAPS, which then relays the data to the intended edge UEs. To design beams at both BSs and HAPS, we formulate an {edge-UE} sum-rate maximization problem under total transmit-power and minimum quality-of-service (QoS) constraints. To solve the resulting non-convex problem, we develop a centralized algorithm based on successive convex approximation (SCA) and alternating optimization (AO) for fast convergence. Simulation results show that relaying information via FD-HAPS significantly improves the capacity of cell-edge UEs compared with a terrestrial-only network.
\end{abstract}

\section{Introduction}
The 3rd Generation Partnership Project (3GPP) Release 17 identifies the integration of terrestrial and non-terrestrial networks (NTNs), including unmanned aerial vehicles (UAVs), high-altitude platform stations (HAPS), and satellites, as a key enabler of ubiquitous connectivity \cite{NBIoT_NTN_Rel_17_2020}.  NTN offers high-probability line-of-sight (LoS) links to ground user equipments (UEs) and wide-area coverage. Among NTN platforms, HAPS has emerged as a key infrastructure for future networks \cite{toka_IEEE_CM_2024, yadav2023IEEEWCM}. 

HAPS, generally solar-powered, operate at 18--22 km altitude and can cover areas up to 200 km in diameter, making them attractive for wide-area communications \cite{ medina-acosta-IEEE-CSM_2026}. Although HAPS were originally introduced to serve remote and sparsely populated regions, this can underutilize their capabilities, especially as metropolitan areas are expected to generate massive and diverse traffic demands \cite{ElJabu2001IEEETVT, Arum2023IEEETMC}. Prior studies have examined both standalone HAPS-based cellular systems and integrated HAPS-terrestrial networks, including learning-based online beamforming \cite{Anjapuli2026VTC}, highlighting HAPS as a promising platform for metropolitan networks \cite{ElJabu2001IEEETVT, Arum2023IEEETMC,yadav2023IEEEWCM}.

Nevertheless, inter-cell interference (ICI) remains a fundamental challenge in terrestrial multicell and next-generation wireless networks, where neighboring base stations (BSs) serving multiple UEs on the same frequency band interfere with one another. This interference is particularly harmful to cell-edge UEs. BS cooperation and coordination with power control can mitigate ICI \cite{3GPP_Report_COMP_2006, gesbert2010IEEEjsac, zhang2010IEEEjsac,3GPP_Report_COMP_2013}. More recently, cell-free multiple-input multiple-output (MIMO) has been proposed to further reduce ICI by allowing multiple distributed BSs to jointly serve all UEs without cell boundaries \cite{nayebi2015asilomar,shaik2020iccworkshops}. Besides ICI, radio-frequency (RF) signals in urban areas suffer from severe shadowing and blockage, which weaken links and often prevent LoS transmission. As a result, BS coordination and cell-free techniques may be less effective under non-LoS (NLoS) and heavy shadowing. Moreover, one HAPS covers many cells, offsetting its cost. 

In this work, we integrate HAPS with terrestrial networks to mitigate ICI, severe shadowing, and blockage, which particularly degrade cell-edge UEs in urban areas.  To the best of our knowledge, this is among the first works to exploit HAPS for improving the capacity of cell-edge UEs in multicell terrestrial networks. Specifically, the terrestrial BSs cooperate with the HAPS, which acts as a relay and forwards their signals to the cell-edge UEs. The proposed algorithm has an antenna-independent per-iteration complexity. HAPS relaying has also been studied for ground-satellite links \cite{HAPS_Relay_2023}.

\begin{figure}[t]
\centerline{\includegraphics[width=0.99\linewidth]{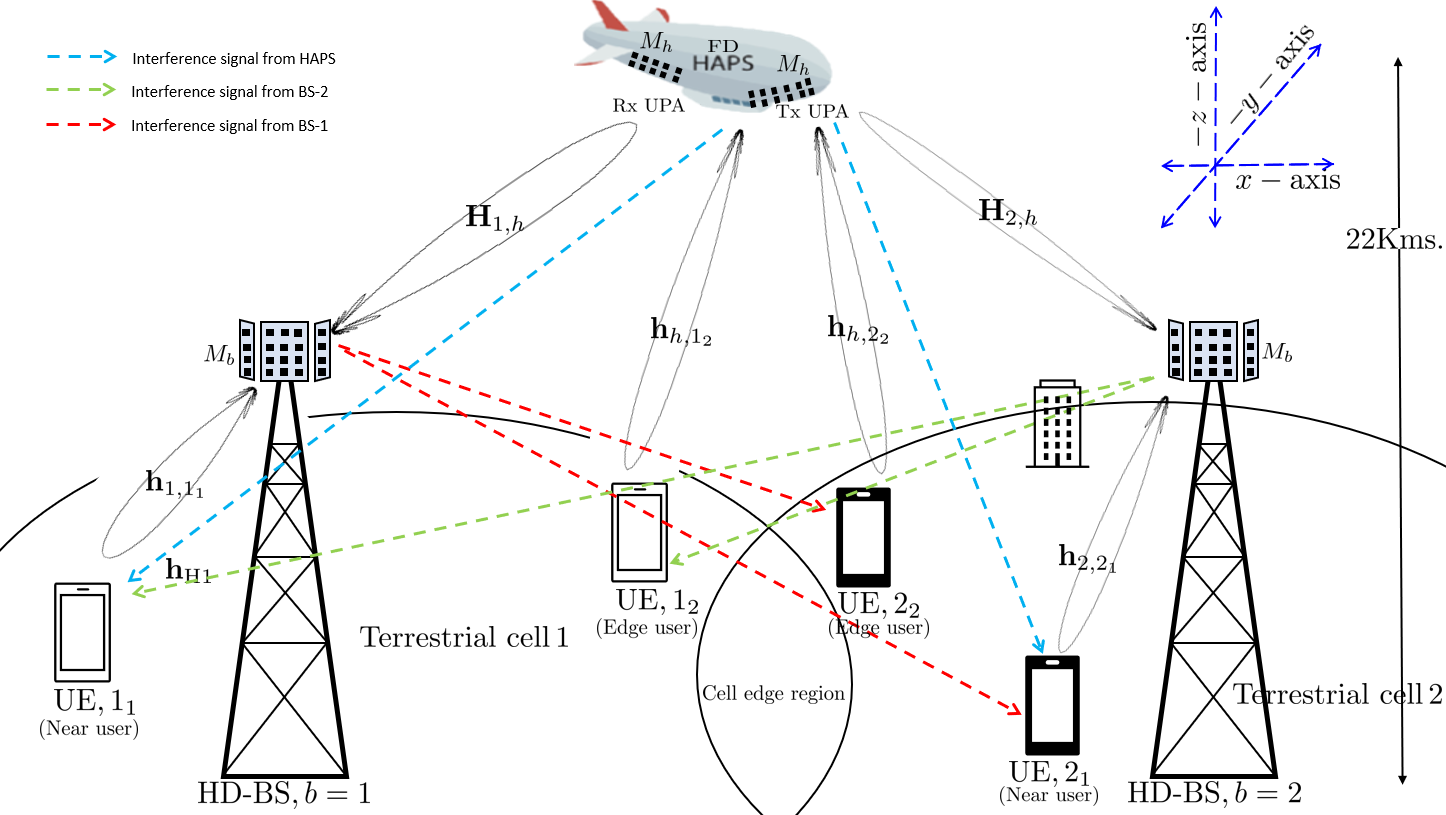}}
\caption{HAPS-complemented terrestrial network. }\label{fig:system_model}
\vspace{-0.1in}
\end{figure}	
\section{System Model}
Consider the HAPS-integrated terrestrial multicell system in Fig.~\ref{fig:system_model}. It comprises $B$ terrestrial cells, each with a multi-antenna BS serving two single-antenna UEs: a near UE $b_1$ and an edge UE $b_2$, for $b=1,\ldots,B$. {Each terrestrial cell has radius $R$; the near UE is located close to its serving BS, whereas the edge UE is located near the cell boundary.} A full-duplex (FD) multi-antenna HAPS (FD-HAPS) at an altitude of 18--22 km, equipped with $M_{\text{h}}$ transmit and $M_{\text{h}}$ receive antennas, covers all cells. Each BS has $M_b$ antennas and operates in conventional half-duplex mode; only the HAPS is FD. The system operates in the millimeter-wave (mmWave) band, where terrestrial links suffer from shadowing and blockage, although large-scale MIMO and beamforming help relax strict LoS requirements.

In urban areas, cell-edge UEs experience lower rates because of severe shadowing, high path loss, and ICI. To mitigate this, the BSs forward edge-UE data via the FD-HAPS, exploiting its LoS links.
 
% \begin{figure}[h]
% \centerline{\includegraphics[width=0.6\linewidth]{angles_plane.PNG}}
% \caption{The angles of elevation and azimuth of an object in a 3D space.}\label{fig:angles_plane}
% \vspace{-0.1in}
% \end{figure}	
\subsection{Channel Model}
We assume that each terrestrial BS uses a uniform planar array (UPA) with $M_b=M_{b_x}M_{b_z}$ antenna elements on the $x$-$z$ plane, where $M_{b_x}$ and $M_{b_z}$ are the numbers of elements along the $x$- and $z$-axes, respectively. The electronically controlled BS radiation orientation is described by the elevation angle $\theta \in [0, \pi]$, measured from the $z$-axis, and the azimuth angle $\varphi \in [0, 2\pi]$. Similarly, the FD-HAPS employs two ground-facing UPAs, one for reception and one for transmission, each with $M_{\text{h}}=M_{\text{h}_x}M_{\text{h}_y}$ elements on the $x$-$y$ plane, where $M_{\text{h}_x}$ and $M_{\text{h}_y}$ are the numbers of elements along the $x$- and $y$-axes, respectively. The antenna spacing and carrier frequency are denoted by $d$ and $f_c$, respectively, and $\lambda=c/f_c$ is the wavelength, where $c$ is the speed of light.

For terrestrial channels, we adopt a sparse geometric multipath channel model with $L$ scatterers, accordingly, the channel vector $\mathbf{h}_{b, b_k}\in \mathbb{C}^{M_b\times 1}$ between the BS $b$ and UE $k$ is given by 
\begin{align}
\mathbf{h}_{b,b_k} = \sum_{i=1}^L\sqrt{\frac{g_b\big(\theta^{t,i}_{b,b_k}, \varphi^{t,i}_{b,b_k}\big)}{L}}\rho^i_{b,b_k}\mathbf{a}_h(\theta^{t,i}_{b,b_k}, \varphi^{t,i}_{b,b_k})\kroneck\mathbf{a}_v(\theta^{t,i}_{b,b_k}),
\end{align}
where $\kroneck$ is Kronecker product operator. $\theta^{t,i}_{b,b_k}(\varphi^{t,i}_{b,b_k})$ denotes the elevation (azimuth) angle of departure (AoD), $g_b\big(\theta^{t,i}_{b,b_k}, \varphi^{t,i}_{b,b_k}\big)$ denotes the antenna gain of BS $b$ and $\rho^i_{b,b_k}$ is the complex channel gain of path $i$, which includes the {NLoS path loss and shadowing (Table~\ref{tab:system_parameter}), as urban BS--UE links are NLoS-dominant~\cite{3GPP_TR38901}}, where $D_{b,b_k}$ is the BS–UE distance. For notational simplicity, superscripts and subscripts are omitted in the horizontal and vertical UPA steering vectors, $\mathbf{a}_h(\theta, \varphi)$ and $\mathbf{a}_v(\theta)$, respectively, given by
\begin{IEEEeqnarray}{lcl}   
    \mathbf{a}_h(\theta, \varphi)&=& \big[1, e^{\frac{2\pi jd}{\lambda}\sin\theta\cos\varphi},\ldots, e^{\frac{2\pi j(M_{b_x}-1)d}{\lambda}\sin\theta\cos\varphi}\big]^T,\IEEEyesnumber\IEEEyessubnumber*\\
    \mathbf{a}_v(\theta) &=&\big[1, e^{\frac{2\pi jd}{\lambda}\cos\theta},\ldots, e^{\frac{2\pi j(M_{b_z}-1)d}{\lambda}\cos\theta}\big]^T.
\end{IEEEeqnarray}

We model aerial channels as a superposition of a dominant LoS component and $L$ NLoS scatterers{ (Rician fading with factor $K$)~\cite{3GPP_TR38811}}. Accordingly, the channel vector $\mathbf{h}_{\text{h},b_k} \in \mathbb{C}^{M_{\text{h}} \times 1}$ between the HAPS and UE $k$ associated with BS $b$ is
\begin{IEEEeqnarray}{l}
\mathbf{h}_{\text{h},b_k} = \sqrt{g_h\big(\theta^{t,0}_{\text{h},b_k}, \varphi^{t,0}_{\text{h},b_k}\big)}\,\rho^0_{\text{h},b_k}\mathbf{a}_x(\theta^{t,0}_{\text{h},b_k}, \varphi^{t,0}_{\text{h},b_k})\!\kroneck\!\mathbf{a}_y(\theta^{t,0}_{\text{h},b_k}, \varphi^{t,0}_{\text{h},b_k}) \nonumber\\
+\sum_{i=1}^L\sqrt{\tfrac{g_h(\theta^{t,i}_{\text{h},b_k}{,}\varphi^{t,i}_{\text{h},b_k})}{L}}\rho^i_{\text{h},b_k}\mathbf{a}_x(\theta^{t,i}_{\text{h},b_k}{,}\varphi^{t,i}_{\text{h},b_k}){\kroneck}\mathbf{a}_y(\theta^{t,i}_{\text{h},b_k}{,}\varphi^{t,i}_{\text{h},b_k}),
\end{IEEEeqnarray}
where $\theta^{t,i}_{\text{h},b_k}(\varphi^{t,i}_{\text{h},b_k})$ denotes the elevation (azimuth) angles, $g_h\big(\theta^{t,0}_{\text{h},b_k}, \varphi^{t,0}_{\text{h},b_k}\big)$ denotes the antenna gain of HAPS and $\rho^i_{\text{h},b_k}$ is the complex channel gain (including the path loss) of path $i$, with $i=0$ representing the LoS path.  For notational simplicity, superscripts and subscripts are omitted in the $x$-axis and $y$-axis steering vectors $\mathbf{a}_x(\theta, \varphi)$ and $\mathbf{a}_y(\theta, \varphi)$, respectively, given by
\begin{IEEEeqnarray}{lcl}
   \mathbf{a}_x(\theta, \varphi) &=& \big[1, e^{\frac{2\pi jd}{\lambda}\sin\theta\cos\varphi},\ldots, e^{\frac{2\pi j(M_{\text{h}_x}-1)d}{\lambda}\sin\theta\cos\varphi}\big]^T,\IEEEyesnumber\IEEEyessubnumber*\\
    \mathbf{a}_y(\theta, \varphi) & = &\big[1, e^{\frac{2\pi jd}{\lambda}\sin\theta\sin\varphi},\ldots, e^{\frac{2\pi j({M_{\text{h}_y}}-1)d}{\lambda}\sin\theta\sin\varphi}\big]^T.
\end{IEEEeqnarray}
Similarly, the channel matrix $\mathbf{H}_{b,\text{h}} \in \mathbb{C}^{M_b\times M_{\text{h}}}$ between BS $b$ and FD-HAPS can be expressed as
\begin{eqnarray}\label{eq:ch_BS_2_HIBS}
\mathbf{H}_{b,\text{h}} = \sqrt{g_h\big(\theta^{r,0}_{b,\text{h}}, \varphi^{r,0}_{b,\text{h}}\big)g_b\big(\theta^{t,0}_{b,\text{h}}, \varphi^{t,0}_{b,\text{h}}\big)}\rho^0_{b,\text{h}}\mathbf{A}_0 \notag\\
+\sqrt{\frac{1}{L}}\sum_{i=1}^L\sqrt{g_h\big(\theta^{r,i}_{b,\text{h}}, \varphi^{r,i}_{b,\text{h}}\big)g_b\big(\theta^{t,i}_{b,\text{h}}, \varphi^{t,i}_{b,\text{h}})}{\rho^i_{b,\text{h}}}\mathbf{A}_i,
\end{eqnarray}
where {$\mathbf{A}_i = \big(\mathbf{a}_h(\theta^{t,i}_{b,\text{h}}, \varphi^{t,i}_{b,\text{h}})\kroneck \mathbf{a}_v(\theta^{t,i}_{b,\text{h}})\big)\big(\mathbf{a}_x(\theta^{r,i}_{b,\text{h}}, \varphi^{r,i}_{b,\text{h}})\kroneck\mathbf{a}_y(\theta^{r,i}_{b,\text{h}}, \varphi^{r,i}_{b,\text{h}})\big)^{H}$}; $\theta^{t,i}\,(\varphi^{t,i})$ and $\theta^{r,i}\,(\varphi^{r,i})$ denote the elevation (azimuth) AoDs and angles of arrival (AoAs), respectively, of path $i$. 

At the FD-HAPS, we ignore the self-interference (SI) channel for two reasons: {at an altitude of 18--22 km}, the SI channel is dominated by the LoS path because reflections are negligible, and {the large separation between the highly directional transmit and receive arrays provides high passive suppression}.

\subsection{Signal Model}
Assume that in transmission time slot $t$, BS $b$ simultaneously serves the near and edge UEs using two beams {on the same time-frequency resource, with successive interference cancellation (SIC) at the near UE as in non-orthogonal multiple access (NOMA)}. The BS is equipped with two RF chains to {generate} two beams and hence, employs a hybrid beamforming architecture for transmission. Specifically, BS $b$ applies a $2\times 2$ baseband precoder $\mathbf{F}^{\text{BB}}_b = [\mathbf{f}^{\text{BB}}_{b_1}, \mathbf{f}^{\text{BB}}_{b_2}]$ followed by an $M_{b}\times 2$ RF precoder $\mathbf{F}^{\text{RF}}_b=[\mathbf{f}^{\text{RF}}_{b_1},\mathbf{f}^{\text{RF}}_{b_2}]$. {The data symbols have unit power, i.e., $\mathbb{E}\{|s_{b_i}[t]|^2\}=1$.} The received signal at the FD-HAPS is
\begin{eqnarray}
&\mathbf{y}_{\text{h}}[t] =& \sum_{b=1}^B\mathbf{H}^H_{b,\text{h}}\sum_{i=1}^{2}\mathbf{F}_b^{\text{RF}}\mathbf{f}_{b_i}^{\text{BB}}s_{b_i}[t] + \mathbf{n}_{\text{h}}[t],
\end{eqnarray}
{where $\mathbf{n}_{\text{h}}[t]\sim\mathcal{CN}(\mathbf{0},\sigma_{\text{h}}^2\mathbf{I}_{M_{\text{h}}})$ is the independent and identically distributed (i.i.d.) additive white Gaussian noise (AWGN) vector at the FD-HAPS.} The FD-HAPS also employs a hybrid combiner with $B$ RF chains to retrieve the received signal from each BS. Let $\mathbf{w}_{b} \triangleq \mathbf{W}_{\text{RF}}\mathbf{w}_{\text{BB},b}$ be the hybrid combiner to retrieve signal from BS $b$. $\mathbf{W}_{\text{RF}} = [\mathbf{w}_{\text{RF},1}, \ldots,\mathbf{w}_{\text{RF},B}]$ is a $M_{\text{h}}\times B$ analog RF {combiner} to process the received signal followed by a $B\times B$ digital baseband {combiner} $\mathbf{W}_{\text{BB}} = [\mathbf{w}_{\text{BB},1}, \ldots, \mathbf{w}_{\text{BB},B}]$.  After the combining process at the HAPS, the signal corresponding to edge UE $b_2,~b\in\{1,\dots, B\}$ is given as
\begin{eqnarray}
r_{\text{h},b_2}[t] &=& \mathbf{w}^H_{b}\bigg({\sum_{m=1}^B\mathbf{H}^H_{m,\text{h}}\sum_{i=1}^{2}\mathbf{F}_m^{\text{RF}}\mathbf{f}_{m_i}^{\text{BB}}s_{m_i}[t]} + \mathbf{n}_{\text{h}}[t]\bigg).\quad
\end{eqnarray}
{The FD-HAPS decodes the edge-UE signals by SIC in increasing order of the BS--HAPS distance, subtracting each decoded signal~\cite{saito2013vtc}. Thus, the signal of BS $b$ is interfered by the near-UE signals of all BSs and the edge-UE signals of the set $\mathcal{U}_b$ of BSs decoded after BS $b$}{, i.e., $\mathcal{U}_b=\{m\neq b: d_{m,\text{h}}>d_{b,\text{h}}\}$, where $d_{m,\text{h}}$ is the distance between BS $m$ and the FD-HAPS.} Therefore, the received signal-to-interference-plus-noise ratio (SINR) of the edge UE $b_2$, corresponding to BS $b$, at the FD-HAPS can be expressed as\predisplaypenalty=0
\begin{equation}
\scalebox{0.9}{$\begin{aligned}\gamma^{\text{h}}_{b_2}={}&|\mathbf{w}^H_{{b}}\mathbf{H}^H_{b,\text{h}}\mathbf{F}^{\text{RF}}_{b}\textbf{f}^{\text{BB}}_{b_2}|^2\Big/\Big(|\mathbf{w}^H_{{b}}\mathbf{H}^H_{b,\text{h}}\mathbf{F}^{\text{RF}}_{b}\textbf{f}^{\text{BB}}_{b_1}|^2+\sigma_{\text{h}}^2\|\mathbf{w}_b\|^2\\&{+\sum\nolimits_{m\neq b}|\mathbf{w}^H_{{b}}\mathbf{H}^H_{m,\text{h}}\mathbf{F}^{\text{RF}}_{m}\textbf{f}^{\text{BB}}_{m_1}|^2+\sum\nolimits_{m\in\mathcal{U}_b}|\mathbf{w}^H_{{b}}\mathbf{H}^H_{m,\text{h}}\mathbf{F}^{\text{RF}}_{m}\textbf{f}^{\text{BB}}_{m_2}|^2}\Big).\end{aligned}$}
\label{eq:dont_use_multline}
\end{equation}

The FD-HAPS, using the NOMA approach, concurrently transmits a composite signal, which is the superimposition of the decoded signals corresponding to the $B$ edge UEs. Assume that the FD-HAPS is also equipped with $B$ RF chains; thus, it employs the hybrid beamforming architecture for transmission \cite{Zhang-Zou-Wang-Ye-Cui-Access-2019}.  In particular, the FD-HAPS applies a $B\times B$  baseband precoder $\mathbf{G}^{\text{BB}} = [\mathbf{g}^{\text{BB}}_{1_2}, \ldots, \mathbf{g}^{\text{BB}}_{B_2}]$ followed by an $M_{\text{h}}\times B$ RF precoder  $\mathbf{G}^{\text{RF}}=[\mathbf{g}^{\text{RF}}_{1_2}, \ldots, \mathbf{g}^{\text{RF}}_{B_2}]$.  The received signal at edge UEs $b_2$ of BS $b$ can be expressed as
\begin{equation}
    r_{b_2}[t] = \mathbf{h}^H_{\text{h},b_2}\sum_{m=1}^B\mathbf{G}^{\text{RF}}\textbf{g}^{\text{BB}}_{m_2}s_{m_2}[t-{\delta}] + {n_{b_2}[t]},
\end{equation}
where {$n_{b_k}[t]\sim\mathcal{CN}(0,\sigma^2_{b_k})$ is the AWGN at UE $b_k$, $k\in\{1,2\}$, and} {$\delta$} is the processing delay at the FD-HAPS \cite{riihonen2011IEEEtsp}.

As the HAPS-to-edge-UE channel gains are similar, each edge UE treats the other edge-UE data as interference, and the SINR of edge UE $b_2$ is given by
\begin{equation}
    \gamma_{b_2}=\frac{|\mathbf{h}^H_{\text{h},b_2}\mathbf{G}^{\text{RF}}\textbf{g}^{\text{BB}}_{b_2}|^2}{\sum_{\substack{m=1 \\ m\neq b}}^B|\mathbf{h}^H_{\text{h},b_2}\mathbf{G}^{\text{RF}}\textbf{g}^{\text{BB}}_{m_2}|^2+\sigma^2_{b_2}}.
\end{equation}

The received signal at near UE $b_1$ is
\begin{IEEEeqnarray}{rCl}
    r_{b_1}[t] &=& \mathbf{h}^H_{b,b_1}\sum_{i=1}^{2}\mathbf{F}_b^{\text{RF}}\mathbf{f}_{b_i}^{\text{BB}}s_{b_i}[t] \nonumber \\
    &&+ \mathbf{h}^H_{\text{h},b_1}\sum_{m=1}^{B}\mathbf{G}^{\text{RF}}\mathbf{g}_{m_2}^{\text{BB}}s_{m_2}[t-{\delta}]
    + {n_{b_1}[t]}.
\end{IEEEeqnarray}
Owing to narrow mmWave beams, high path loss, and the proximity of near UEs to their serving BS, interference from other terrestrial BSs at near UEs is omitted; likewise, terrestrial signals at edge UEs are neglected due to heavy shadowing. The near UE $b_1$ applies SIC by decoding and subtracting the edge-UE data, including the HAPS-relayed $s_{b_2}[t-\delta]$ decoded in slot $t-\delta$. Thus, the SINR of near UE $b_1$ is
\begin{flalign}
    \gamma_{b_1} = \frac{|\mathbf{h}^H_{b,b_1}\mathbf{F}_b^{\text{RF}}\mathbf{f}_{b_1}^{\text{BB}}|^2}{{\sum_{m\neq b}}|\mathbf{h}^H_{\text{h},b_1}\mathbf{G}^{\text{RF}}\mathbf{g}_{m_2}^{\text{BB}}|^2+\sigma^2_{b_1}},
\end{flalign}
and the received SINR of edge UE $b_2$ at near UE $b_1$ can be expressed as
\begin{flalign}
    \gamma_{b_{1\rightarrow 2}}= \frac{|\mathbf{h}^H_{b,b_1}\mathbf{F}_b^{\text{RF}}\mathbf{f}_{b_2}^{\text{BB}}|^2}{|\mathbf{h}^H_{b,b_1}\mathbf{F}_b^{\text{RF}}\mathbf{f}_{b_1}^{\text{BB}}|^2 + {\sum_{m\neq b}}|\mathbf{h}^H_{\text{h},b_1}\mathbf{G}^{\text{RF}}\mathbf{g}_{m_2}^{\text{BB}}|^2+\sigma^2_{b_1}}.
\end{flalign}
The achievable rates for the near and edge UEs in each cluster corresponding to each BS $b \in \{1, \ldots, B\}$ can be expressed, respectively, as
\begin{IEEEeqnarray}{lcl}\label{eq:rate_edge_UEs}
R_{b_1} = \log_2(1+\gamma_{b_1}),\,\, R_{b_2} = \log_2(1+\min\{\gamma_{b_{1\rightarrow 2}}, \gamma^{\text{h}}_{b_2}, \gamma_{b_2}\}).\IEEEyesnumber \IEEEeqnarraynumspace 
\end{IEEEeqnarray}
Note that in \eqref{eq:rate_edge_UEs}, $\gamma_{b_{1\rightarrow 2}}$ is included to ensure that near UEs $b_1$ are successfully able to perform SIC.

\section{Problem Formulation}
In this section, we formulate a joint optimization problem for the BS and FD-HAPS hybrid precoders and the FD-HAPS receive combiner to maximize the {edge-UE} sum rate, subject to transmit-power and minimum-rate quality-of-service (QoS) $(r_{\text{QoS}})$ constraints for {all UEs}. The resulting problem is expressed as
\begin{IEEEeqnarray*}{lcl}\label{eq:Orig_P1}
&\underset{\substack{\mathcal{F}^{\text{RF}}, \mathcal{F}^{\text{BB}}, \mathbf{G}^{\text{RF}},\\ \mathbf{G}^{\text{BB}}, \mathbf{W}_{\text{RF}}, \mathbf{W}_{\text{BB}}}}{\max}\,\, & {\sum_{b=1}^B R_{b_2}} \IEEEyesnumber \IEEEyessubnumber* \label{eq:P1_Obj}\\
&\text{s.t.} & {R_{b_k}}\geq r_{\text{QoS}}, \forall b{,k}, \label{eq:P1_far_UE_rate_constr}\\
&& \abs{[\mathbf{F}_b^{\text{RF}}]_{i,j}}^2 = 1/M_b,\, \forall i, j, b, \label{eq:P1_BS_const_mod_constr}\\
&& \abs{[\mathbf{G}^{\text{RF}}]_{i,j}}^2 = 1/M_{\text{h}}, \, \forall i, j, \label{eq:P1_HIBS_const_mod_constr}\\
&& \abs{[{\mathbf{w}_{\text{RF},b}}]_{i}}^2 = 1/M_{\text{h}},\, \forall i,b,\label{eq:P1_HIBS_combiner_const_mod_constr}\\
&& \textstyle\sum_{i=1}^2\norm{\mathbf{F}_b^{\text{RF}}\mathbf{f}_{b_i}^{\text{BB}}}_F^2 \leq P_{\text{max},b},\, \forall b, \label{eq:P1_BS_mat_product_constr}\\
&& {\textstyle\sum_{m=1}^B\norm{\mathbf{G}^{\text{RF}}\mathbf{g}_{m_2}^{\text{BB}}}_F^2} \leq P_{\text{max, h}},\label{eq:P1_HIBS_mat_product_constr}
\end{IEEEeqnarray*}
where $\mathcal{F}^{\text{RF}}=[\mathbf{F}^{\text{RF}}_1,\ldots,\mathbf{F}^{\text{RF}}_B]$ and $\mathcal{F}^{\text{BB}}=[\mathbf{f}^{\text{BB}}_{1_1},\mathbf{f}^{\text{BB}}_{1_2},\ldots,\mathbf{f}^{\text{BB}}_{B_1},\mathbf{f}^{\text{BB}}_{B_2}]$ collect the RF and baseband precoders for BSs $b\in\{1,\ldots,B\}$, respectively. $P_{\text{max},b}$ and $P_{\text{max, h}}$ denote the maximum transmit powers of each BS and the FD-HAPS, respectively. Constraint \eqref{eq:P1_far_UE_rate_constr} guarantees a minimum rate of $r_{\text{QoS}}$ bits/s/Hz {for every UE}. Constraints \eqref{eq:P1_BS_const_mod_constr}--\eqref{eq:P1_HIBS_combiner_const_mod_constr} impose constant-modulus RF precoding/combining with magnitudes $M_b^{-1/2}$, $M_{\text{h}}^{-1/2}$, and $M_{\text{h}}^{-1/2}$, respectively, while \eqref{eq:P1_BS_mat_product_constr} and \eqref{eq:P1_HIBS_mat_product_constr} enforce the BS and FD-HAPS transmit-power limits. Further, for coordination, one BS acts as a centralized server that collects global information from the other {BSs} and the HAPS via a dedicated X2 interface and wireless link, respectively, assuming perfect channel state information (CSI).

\section{Proposed Solution}
Problem \eqref{eq:Orig_P1} is non-convex because both the objective function \eqref{eq:P1_Obj} and the feasible set defined by constraints \eqref{eq:P1_far_UE_rate_constr}--\eqref{eq:P1_HIBS_combiner_const_mod_constr} are non-convex, making the global optimum difficult to obtain. Therefore, we adopt a simplified RF precoder design in which each phase is set using the phase of the conjugate channel. Accordingly, the RF precoder for the near UE of BS $b$ is
\begin{eqnarray}
\mathbf{f}^{\text{RF}}_{b_1} = {\tfrac{1}{\sqrt{M_b}}}[e^{-j\angle [\mathbf{h}^{*}_{b,b_1}]_1}, \ldots, e^{-j\angle [\mathbf{h}^{*}_{b,b_1}]_{M_b}}]^T,\, \forall b, \label{eq:RF_precoder_near_UEs}
\end{eqnarray}
where $[\mathbf{h}_{b,b_1}]_i$ is the $i$th element of the channel vector $\mathbf{h}_{b,b_1}$. And the RF precoder corresponding to the edge UEs of BS $b$ can be written as $\mathbf{f}^{\text{RF}}_{b_2} = {\tfrac{1}{\sqrt{M_b}}}\mathbf{a}_t(\theta^{t,0}_{b,\text{h}}, \varphi^{t,0}_{b,\text{h}}),\, \forall b${, where $\mathbf{a}_t(\theta,\varphi)\triangleq\mathbf{a}_h(\theta,\varphi)\kroneck\mathbf{a}_v(\theta)$}.
% \begin{eqnarray}
% \mathbf{f}^{\text{RF}}_{b_2} = \mathbf{a}_t(\theta^{t,0}_{b,\text{h}}, \varphi^{t,0}_{b,\text{h}}),\quad \forall b. \label{eq:RF_precoder_near_UEs}
% \end{eqnarray}
Next, the RF precoder corresponding to the edge UEs at the HAPS can be written as
\begin{eqnarray}
\mathbf{g}^{\text{RF}}_{b_2} = {\tfrac{1}{\sqrt{M_{\text{h}}}}}[e^{-j\angle [\mathbf{h}^{*}_{\text{h},b_2}]_1}, \ldots, e^{-j\angle [{\mathbf{h}^{*}_{\text{h},b_2}}]_{M_{\text{h}}}}]^T,\, \forall b,\label{eq:RF_precoder_far_UEs}
\end{eqnarray}
where {$[\mathbf{h}_{\text{h},b_2}]_i$} is element $i$ of the channel vector $\mathbf{h}_{\text{h},b_2}$.
{Similarly, the RF combiner vector corresponding to the signal received from BS $b$ at the FD-HAPS can be obtained as $\mathbf{w}_{\text{RF},b} = {\tfrac{1}{\sqrt{M_{\text{h}}}}}\mathbf{a}_r(\theta^{r,0}_{b,\text{h}}, \varphi^{r,0}_{b,\text{h}}),\, \forall b$, where $\mathbf{a}_r(\theta,\varphi)\triangleq\mathbf{a}_x(\theta,\varphi)\kroneck\mathbf{a}_y(\theta,\varphi)$.}
% \begin{eqnarray}
% \mathbf{w}^{\text{RF}}_{\text{h}, b_2} = \mathbf{a}_r(\theta^{t,0}_{b,\text{h}}, \varphi^{t,0}_{b,\text{h}}),\quad \forall b. \label{eq:RF_precoder_near_UEs}
% \end{eqnarray}

With the RF precoders $\mathbf{F}^{\text{RF}}_b$ and $\mathbf{G}^{\text{RF}}$, and RF combiner {$\mathbf{W}_{\text{RF}}$} acquired as above, we now define the effective channels between BS $b$ and near UEs, and between HAPS near and edge UEs, respectively, as
\begin{eqnarray}
\tilde{\mathbf{h}}_{b, b_1} = \mathbf{h}^H_{b, b_1}\mathbf{F}^{\text{RF}}_b, \forall b, \text{  and   } \tilde{\mathbf{h}}_{\text{h}, b_1} = \mathbf{h}^H_{\text{h}, b_1}\mathbf{G}^{\text{RF}}, \forall b,\label{eq:effe_channels}
\end{eqnarray}
\begin{eqnarray}
\tilde{\mathbf{h}}_{\text{h}, b_2} = \mathbf{h}^H_{\text{h}, b_2}\mathbf{G}^{\text{RF}}, \forall b, \text{  and  }\tilde{\mathbf{H}}_{b,\text{h}} = \mathbf{W}_{\text{RF}}^H\mathbf{H}^H_{b,\text{h}}\mathbf{F}^{\text{RF}}_{b},\, \forall b.
\end{eqnarray}

With fixed RF precoders, \eqref{eq:Orig_P1} is reduced to
\vspace{-0.05 in}
\begin{IEEEeqnarray*}{lcl}\label{eq:P2}
&\underset{\mathcal{F}^{\text{BB}}, \mathbf{G}^{\text{BB}}, \mathbf{W}_{\text{BB}}}{\text{maximize}}\,\, & {\sum_{b=1}^B R_{b_2}} \IEEEyesnumber \IEEEyessubnumber* \label{eq:P2_Obj}\\
&\text{s.t.} & \eqref{eq:P1_far_UE_rate_constr},{\eqref{eq:P1_BS_mat_product_constr},\eqref{eq:P1_HIBS_mat_product_constr}}.
\end{IEEEeqnarray*}
Next, we transform \eqref{eq:P2} to expose hidden convexity and approximate the remaining non-convex constraints, yielding a fast, low-complexity successive convex approximation (SCA)-based iterative algorithm \cite{beck2010jgo}.
\subsection{Equivalent Transformations}
Introducing the slack variables $\tau_{b_2}$, $\forall\,b$, as lower bounds on the three SINRs, tight at the optimum, we rewrite \eqref{eq:P2} equivalently as
\vspace{-0.05 in}
\begin{IEEEeqnarray*}{lcl}\label{eq:P3}
&\underset{\substack{\mathcal{F}^{\text{BB}}, \mathbf{G}^{\text{BB}}, \mathbf{W}_{\text{BB}}, \bm{\tau}}}{\text{maximize}}\,\quad & {\sum_{b=1}^B\log_2(1+\tau_{b_2})} \IEEEyesnumber \IEEEyessubnumber* \IEEEeqnarraynumspace\label{eq:P3_Obj}\\
&\text{s.t.} & \gamma_{b_{1\rightarrow 2}} \geq \tau_{b_2}, \, \forall b,\label{eq:P3_const1}\\
&& \gamma^{\text{h}}_{b_{2}} \geq \tau_{b_2}, \, \forall b,\label{eq:P3_const2}\\
&& \gamma_{b_{2}} \geq \tau_{b_2}, \, \forall b,\label{eq:P3_const3}\\
&& \tau_{b_2} \geq \Gamma_{b_2}^{\text{min}}, \, \forall b, \label{eq:P3_const4}\\
&& \gamma_{b_1}\geq \Gamma_{b_1}^{\text{min}}, \, \forall b,\label{eq:P3_const5}\\
&&\eqref{eq:P1_BS_mat_product_constr}-\eqref{eq:P1_HIBS_mat_product_constr},\label{eq:P3_const6}
\end{IEEEeqnarray*}
where $\Gamma_{b_{k}}^{\text{min}} = 2^{r_{\text{QoS}}}-1$ and $\bm{\tau}=\{\tau_{1_2},\ldots, \tau_{B_2}\}$.
Writing \eqref{eq:P3_Obj} as ${\log_2\prod_{b=1}^B (1+\tau_{b_2})}$ and noticing that the logarithm is monotonically increasing, \eqref{eq:P3} is equivalent to
% \begin{IEEEeqnarray*}{lcl}\label{eq:P4}
% &\underset{\substack{\mathcal{F}^{\text{BB}}, \mathbf{G}^{\text{BB}},\\ \mathbf{W}_{\text{BB}}, \bm{\tau}}}{\text{maximize}}\,\, & \prod_{b=1}^B (1+\gamma_{b_1})(1+\tau_{b_2}) \IEEEyesnumber \IEEEyessubnumber* \label{eq:P4_Obj}\\
% &\text{s.t.} & \eqref{eq:P3_const1}-\eqref{eq:P3_const4}, \eqref{eq:P1_BS_mat_product_constr}-\eqref{eq:P1_HIBS_mat_product_constr},\label{eq:P4_const1}
% \end{IEEEeqnarray*}
% which can be equivalently expressed as
\vspace{-0.05 in}
\begin{IEEEeqnarray*}{lcl}\label{eq:P5}
&\underset{\substack{\mathcal{F}^{\text{BB}}, \mathbf{G}^{\text{BB}}, \mathbf{W}_{\text{BB}}, \bm{\tau}, \mathbf{z}}}{\text{maximize}}\,\quad & {\prod_{b=1}^B z_{b_2}} \IEEEyesnumber \IEEEyessubnumber* \label{eq:P5_Obj}\\
&\text{s.t.} & \gamma_{b_1} \geq z_{b_1}-1, \, \forall b,\label{eq:P5_const1}\\
&& \tau_{b_2} \geq z_{b_2}-1, \, \forall b,\label{eq:P5_const2}\\
&& z_{b_1}\geq 1+\Gamma_{b_1}^{\text{min}}, \, \forall b,\label{eq:P5_const2b}\\
&&\eqref{eq:P3_const1}-\eqref{eq:P3_const4}, \eqref{eq:P1_BS_mat_product_constr}-\eqref{eq:P1_HIBS_mat_product_constr},\label{eq:P5_const3}
\end{IEEEeqnarray*}
where $\mathbf{z}$ collects variables $\{z_{1_1}, z_{1_2}, \ldots, z_{B_1}, z_{B_2}\}${, and $\mathbf{z}_2=\{z_{1_2},\ldots,z_{B_2}\}$}, where $z_{b_k}$ represents $1+\text{SINR}$ of UE $b_k$.
Next, by introducing $B$ new additional slack variables $\beta_{b_1}, \forall b \in \{1, \ldots, B\}$, which upper-bound the interference-plus-noise power, constraint \eqref{eq:P5_const1} can be decomposed into the following pair of inequalities as
\begin{IEEEeqnarray*}{lcl}\label{eq:P5_const1_ab}
\vspace{-0.05 in}
\abs{\tilde{\mathbf{h}}_{b,b_1}\mathbf{f}_{b_1}^{\text{BB}}}^2 \geq (z_{b_1}-1)\beta_{b_1}, \, \forall b,\IEEEyesnumber \IEEEyessubnumber* \label{eq:P5_const1_a}\\
{\sum\nolimits_{m\neq b}\abs{\tilde{\mathbf{h}}_{\text{h},b_1}\mathbf{g}_{m_2}^{\text{BB}}}^2}+\sigma^2_{b_1} \leq \beta_{b_1}, \, \forall b{.}\label{eq:P5_const1_b}
\end{IEEEeqnarray*}
Similarly, by introducing additional slack variables $\{\beta_{b_2}, \ldots, \beta_{B_2}, \beta^{\text{h}}_{1_2}, \ldots, \beta^{\text{h}}_{B_2}, \beta_{1_{1\rightarrow 2}}, \ldots,\beta_{B_{1\rightarrow 2}}\}$, constraints \eqref{eq:P3_const1}, \eqref{eq:P3_const2}, and \eqref{eq:P3_const3} can be decomposed equivalently into the following pair of inequalities, respectively, as
\vspace{-0.05 in}
\begin{IEEEeqnarray*}{c}\label{eq:P3_const1_ab}
\abs{\tilde{\mathbf{h}}_{b,b_1}\mathbf{f}_{b_2}^{\text{BB}}}^2  \geq \tau_{b_2}\beta_{b_{1\rightarrow 2}}{,} \, \forall b, \IEEEyesnumber \IEEEyessubnumber* \label{eq:P3_const1_a} \\
\abs{\tilde{\mathbf{h}}_{b,b_1}\mathbf{f}_{b_1}^{\text{BB}}}^2 + {\sum\nolimits_{m\neq b}\abs{\tilde{\mathbf{h}}_{\text{h},b_1}\mathbf{g}_{m_2}^{\text{BB}}}^2}+\sigma^2_{b_1} \leq \beta_{b_{1\rightarrow 2}}, \, \forall b, \label{eq:P3_const1_b}
\end{IEEEeqnarray*}
\begin{IEEEeqnarray*}{rCl}
\IEEEeqnarraymulticol{3}{c}{|{\mathbf{w}^H_{\text{BB},b}\tilde{\mathbf{H}}_{b,\text{h}}}\textbf{f}^{\text{BB}}_{b_2}|^2 \geq \tau_{b_2}\beta^{\text{h}}_{b_2}, \, \forall b,}\IEEEyesnumber \IEEEyessubnumber*\label{eq:P3_const2_ab}\label{eq:P3_const2_a}\\
&&|{\mathbf{w}^H_{\text{BB},b}\tilde{\mathbf{H}}_{b,\text{h}}}\textbf{f}^{\text{BB}}_{b_1}|^2+\sigma_{\text{h}}^2\|\mathbf{W}_{\text{RF}}\mathbf{w}_{\text{BB},b}\|^2{+\sum\nolimits_{m\neq b}|{\mathbf{w}^H_{\text{BB},b}\tilde{\mathbf{H}}_{m,\text{h}}}\textbf{f}^{\text{BB}}_{m_1}|^2} \nonumber\\
&&{+\sum\nolimits_{m\in\mathcal{U}_b}|{\mathbf{w}^H_{\text{BB},b}\tilde{\mathbf{H}}_{m,\text{h}}}\textbf{f}^{\text{BB}}_{m_2}|^2} \leq \beta^{\text{h}}_{b_2},\, \forall b, \label{eq:P3_const2_b}
\end{IEEEeqnarray*}
\begin{IEEEeqnarray*}{c}\label{eq:P3_const31_ab}
\abs{\tilde{\mathbf{h}}_{\text{h},b_2}\textbf{g}^{\text{BB}}_{b_2}}^2 \geq \tau_{b_2}\beta_{b_2}, \, \forall b,\IEEEyesnumber \IEEEyessubnumber*\label{eq:P3_const31_a}
\end{IEEEeqnarray*}
\begin{IEEEeqnarray*}{c}
\sum_{m=1, m\neq b}^B\abs{\tilde{\mathbf{h}}_{\text{h},b_2}\textbf{g}^{\text{BB}}_{m_2}}^2+\sigma^2_{b_2} \leq \beta_{b_{2}}, \, \forall b. \IEEEyessubnumber*\label{eq:P3_const31_b}
\end{IEEEeqnarray*}
Using constraints \eqref{eq:P5_const1_ab}-\eqref{eq:P3_const31_ab}, problem \eqref{eq:P5} can be reformulated as 
\vspace{-0.05 in}
\begin{IEEEeqnarray*}{lcl}\label{eq:P6}
&\underset{\substack{\mathcal{F}^{\text{BB}}, \mathbf{G}^{\text{BB}}, \mathbf{W}_{\text{BB}}, \bm{\tau}, \mathbf{z},\bm{\beta}}}{\text{maximize}}\,\quad & \texttt{geo\_mean}({\mathbf{z}_2}) \IEEEyesnumber \IEEEyessubnumber* \label{eq:P6_Obj}\\
&\text{s.t.} & \eqref{eq:P1_BS_mat_product_constr}\text{--}\eqref{eq:P1_HIBS_mat_product_constr}, \eqref{eq:P3_const4}, \eqref{eq:P5_const2}\text{--}\eqref{eq:P5_const2b}, \nonumber\\
&& \eqref{eq:P5_const1_ab}\text{--}\eqref{eq:P3_const31_ab},
\end{IEEEeqnarray*}
where, $\texttt{geo\_mean}(\mathbf{z})$ denotes the geometric mean of $\mathbf{z}$ and is concave and increasing. Problem \eqref{eq:P6} remains non-convex due to constraints \eqref{eq:P5_const1_a}, \eqref{eq:P3_const1_a}, \eqref{eq:P3_const2_a}, and \eqref{eq:P3_const31_a}. We therefore adopt an alternating optimization (AO) technique. This technique decomposes the problem into two tractable subproblems over set of variables $\{\mathcal{F}^{\text{BB}}, \mathbf{G}^{\text{BB}}\}$ and $\mathbf{W}_{\text{BB}}$.

\subsection{Problem Approximations}
For a fixed baseband combiner $\mathbf{W}_{\text{BB}}$, problem \eqref{eq:P6} reduces to optimizing the transmit baseband precoders $\mathcal{F}^{\text{BB}}$ and $\mathbf{G}^{\text{BB}}$, which remains challenging. {In this case, constraints} \eqref{eq:P5_const1_b}, \eqref{eq:P3_const1_b}, \eqref{eq:P3_const2_b}, and \eqref{eq:P3_const31_b} can be written as second-order cone (SOC) constraints. The remaining difficulty lies in the non-convex constraints \eqref{eq:P5_const1_a}, \eqref{eq:P3_const1_a}, \eqref{eq:P3_const2_a}, and \eqref{eq:P3_const31_a}, where the convex function $f(x,y)\triangleq \abs{x}^2/y$, for $x\in\mathbb{C}$ and $y\in\mathbb{R}$, appears on the upper side of the inequality. We approximate such terms using a first-order Taylor expansion around $(x^{(n)}, y^{(n)})$ at the $n$th iteration. Let the real and imaginary parts of $x\in \mathbb{C}$ be
\begin{eqnarray}
    p \triangleq \Re\{x\}
    \mbox{  and  }q \triangleq \Im\{x\},
\end{eqnarray}
such that $f(x,y) = (p^2+q^2)/y$. $\Re\{\cdot\}$ and $\Im\{\cdot\}$ denote the real and imaginary parts of a complex scalar, respectively. Now, we are in position to write the first order Taylor approximation of $f(x,y)$ around a fixed point $(x^{(n)}, y^{(n)})$ at the $n$th iteration as
\begin{IEEEeqnarray}{cRl}
    F^{(n)}(x,y{;\,x^{(n)},y^{(n)}}) &=& 2\frac{p^{(n)}}{y^{(n)}}\big(p-p^{(n)}\big) + 2\frac{q^{(n)}}{y^{(n)}}\big(q-q^{(n)}\big)\nonumber\\
&& \hspace{-1.5em} + \frac{(p^{(n)})^2 + (q^{(n)})^2}{y^{(n)}}\bigg(1-\frac{y-y^{(n)}}{y^{(n)}}\bigg),\label{eq:Taylor_F}
\end{IEEEeqnarray}

Therefore, for fixed $\mathbf{W}_{\text{BB}}$, the transmit baseband precoders are obtained by iteratively solving the following relaxed convex subproblem {at the $n$th SCA iteration}:
\begin{IEEEeqnarray*}{lcl}\label{eq:P6_sub1}
\hspace*{-2.5em}&\underset{\Xi}{\text{maximize}}\,\quad & {\texttt{geo\_mean}({\mathbf{z}_2})} \IEEEyesnumber \IEEEyessubnumber* \label{eq:P6_sub1_Obj}\\
\hspace*{-2.5em}&\text{s.t.} & F_1^{(n)}(\mathbf{f}_{b_1}^{\text{BB}}, \beta_{b_1}{;\,\mathbf{f}_{b_1}^{\text{BB},(n)},\beta_{b_1}^{(n)}}) \geq z_{b_1}-1,{\ \forall b},\quad \label{eq:P6_sub1_F1}\\
\hspace*{-2.5em}&& F_2^{(n)}(\mathbf{f}_{b_2}^{\text{BB}}, \beta_{b_{1 \rightarrow 2}}{;\,\mathbf{f}_{b_2}^{\text{BB},(n)},\beta_{b_{1\rightarrow 2}}^{(n)}}) \geq \tau_{b_2},{\ \forall b},\\
\hspace*{-2.5em}&& {F_3^{(n)}(\mathbf{f}_{b_2}^{\text{BB}}, \beta^{\text{h}}_{b_2}{;\,\mathbf{f}_{b_2}^{\text{BB},(n)},\beta^{\text{h},(n)}_{b_2}}) \geq \tau_{b_2},{\ \forall b},}\\
\hspace*{-2.5em}&& {F_4^{(n)}}(\mathbf{g}_{b_2}^{\text{BB}}, \beta_{b_2}{;\,\mathbf{g}_{b_2}^{\text{BB},(n)},\beta_{b_2}^{(n)}}) \geq \tau_{b_2},{\ \forall b},\label{eq:P6_sub1_F4}\\
\hspace*{-2.5em}&&\eqref{eq:P1_BS_mat_product_constr}\text{--}\eqref{eq:P1_HIBS_mat_product_constr}, \eqref{eq:P3_const4}, \eqref{eq:P5_const2}\text{--}\eqref{eq:P5_const2b}, \eqref{eq:P5_const1_b}\text{--}\eqref{eq:P3_const31_b},
\end{IEEEeqnarray*}
where $\Xi = \{\mathcal{F}^{\text{BB}}, \mathbf{G}^{\text{BB}}, \bm{\tau}, \mathbf{z},\bm{\beta}\}$ denotes a set that collects all the optimization variable for subproblem \eqref{eq:P6_sub1}. {{The} left-hand sides of \eqref{eq:P6_sub1_F1}--\eqref{eq:P6_sub1_F4} are the Taylor approximations \eqref{eq:Taylor_F} of $f(x,y)$ in \eqref{eq:P5_const1_a}--\eqref{eq:P3_const31_a}, respectively, (e.g., in $F_1^{(n)}$, $x=\tilde{\mathbf{h}}_{b,b_1}\mathbf{f}^{\text{BB}}_{b_1}$ and $y=\beta_{b_1}$, and so on.)} 

Next, the baseband combiner at the FD-HAPS {for fixed transmit baseband precoders is} obtained by solving the following subproblem {at the $n$th SCA iteration}:
\begin{IEEEeqnarray*}{lcl}\label{eq:P6_sub2}
\hspace*{-2.5em}&\underset{\Upsilon}{\text{maximize}}\,\quad & \texttt{geo\_mean}({\mathbf{z}_2}) \IEEEyesnumber \IEEEyessubnumber* \label{eq:P6_sub2_Obj}\\
\hspace*{-2.5em}&\text{s.t.} & G_b^{(n)}({\mathbf{w}_{\text{BB},b}}, \beta^{\text{h}}_{b_2}{;\,\mathbf{w}_{\text{BB},b}^{(n)},\beta^{\text{h},(n)}_{b_2}}) \geq \tau_{b_2},{\ \forall b},\quad \label{eq:P6_sub2_G}\\
\hspace*{-2.5em}&&\eqref{eq:P3_const4}, \eqref{eq:P5_const2}\text{--}\eqref{eq:P5_const2b}, \eqref{eq:P3_const2_b}, {\eqref{eq:P5_const1}, \eqref{eq:P3_const1}, \eqref{eq:P3_const3}},
\end{IEEEeqnarray*}
where the set $\Upsilon = \{\mathbf{W}_{\text{BB}}, \bm{\tau}, \mathbf{z},\bm{\beta}\}$ collects all the optimization variables, and the function $G_b^{(n)}({\mathbf{w}_{\text{BB},b}}, \beta^{\text{h}}_{b_2}{;\,\mathbf{w}_{\text{BB},b}^{(n)},\beta^{\text{h},(n)}_{b_2}})$ represents the {first-order Taylor approximation of $f(x,y)$ in constraint \eqref{eq:P3_const2_a}} corresponding to BS $b$. {Since $\gamma_{b_1}$, $\gamma_{b_{1\rightarrow 2}}$, and $\gamma_{b_2}$ do not depend on $\mathbf{W}_{\text{BB}}$, they are evaluated at the fixed precoders, so that \eqref{eq:P5_const1}, \eqref{eq:P3_const1}, and \eqref{eq:P3_const3} are linear in $\mathbf{z}$ and $\bm{\tau}$.}

\begin{algorithm}[t]
\caption{Iterative AO algorithm}\label{algo:Algo1}
\footnotesize
\algrenewcommand\algorithmicrequire{\textbf{Input:}}
\algrenewcommand\algorithmicensure{\textbf{Output:}}
\begin{algorithmic}[1]
\Require $r_{\text{QoS}}$, $\tilde{\mathbf{h}}_{b,b_1}$, $\tilde{\mathbf{h}}_{\text{h},b_1}$, $\tilde{\mathbf{h}}_{\text{h},b_2}$, $\tilde{\mathbf{H}}_{b,\text{h}}$, $P_{\text{max},b}$, {$\sigma^2_{b_1}$, $\sigma^2_{b_2}$}, $\forall b$; $P_{\text{max},\text{h}}$, {$\sigma^2_{\text{h}}$, $\mathbf{W}_{\text{RF}}$}, $I_{\text{max}}$
\Ensure {$\mathcal{F}^{\text{BB}}$}, $\mathbf{G}^{\text{BB}}$, {$\mathbf{W}_{\text{BB}}$}
\State Initialize $k:=1$ and {$\mathbf{W}^{(0)}_{\text{BB}}$}
\Repeat
\State {Set $n:=0$} and initialize $\Xi^{(0)}$
\Repeat
\State Solve \eqref{eq:P6_sub1} {around $\Xi^{(n)}$} using {$\mathbf{W}^{(k-1)}_{\text{BB}}$}
\State Update {$\Xi^{(n+1)}=\Xi^{\star}$} and {$n:=n+1$}
\Until{SCA convergence or {$n \geq I_{\text{max}}$}}
\State {Get $\mathcal{F}^{\text{BB},(k)}, \mathbf{G}^{\text{BB},(k)}$ from $\Xi^{(n)}$; set $n:=0$} and initialize $\Upsilon^{(0)}$
\Repeat
\State Solve \eqref{eq:P6_sub2} {around $\Upsilon^{(n)}$} using {$\mathcal{F}^{\text{BB},(k)}$, $\mathbf{G}^{\text{BB},(k)}$}
\State Update {$\Upsilon^{(n+1)}=\Upsilon^{\star}$} and {$n:=n+1$}
\Until{SCA convergence or {$n \geq I_{\text{max}}$}}
\State {Get $\mathbf{W}^{(k)}_{\text{BB}}$ from $\Upsilon^{(n)}$;} $k:=k+1$
\Until{{objective} convergence or {$k > I_{\text{max}}$}}
\end{algorithmic}
\end{algorithm}

% \begin{algorithm}[h]
%   \KwInput{$r_{\text{QoS}}, \tilde{\mathbf{h}}_{b,b_1}, \forall b, \tilde{\mathbf{h}}_{\text{h},b_1},\forall b, \tilde{\mathbf{h}}_{\text{h},b_2},\forall b,  \tilde{\mathbf{H}}_{b,\text{h}}, \forall b, M_{\text{h}},$\\ $P_{\text{max},\text{h}},P_{\text{max},b},\forall b , I_{\text{max}}$} 
%   \KwOutput{$\mathbf{F}^{\text{BB}}_1,\mathbf{F}^{\text{BB}}_2, \mathbf{G}^{\text{BB}}, \mathbf{w}_{\text{BB}}$}
%   \KwData{$k:=1, \mathbf{w}^{(0)}_{\text{BB}}$  }
  
%   \Repeat{$\text{Sum rate convergence or } k\leq I_{\text{max}}$}
%      {set $i:=1$ and initialize $\Xi^{(0)}$,\\
%     \Repeat{$\text{SCA convergence or } i\leq I_{\text{max}}$}
%         {   
%         {solve \eqref{eq:P6_sub1} for transmit baseband precoders $\mathbf{F}^{\text{BB}(k)}_1,\,\mathbf{F}^{\text{BB}(k)}_2, \,\mathbf{G}^{\text{BB}(k)}$ using $\mathbf{w}^{(k-1)}_{\text{BB}}$ }\\
%         {update $\Xi^{(i)}=\Xi^{*}$ and $i = i+1$}
%         }
%         { set $j:=1$ and initialize $\Upsilon^{(0)}$}\\
%         \Repeat{$\text{SCA convergence or } j\leq I_{\text{max}}$}
%         {   
%         {solve \eqref{eq:P6_sub2} for baseband combiner $\mathbf{w}^{(k)}_{\text{BB}}$ using updated $\mathbf{F}^{\text{BB}(k)}_1,\,\mathbf{F}^{\text{BB}(k)}_2, \,\mathbf{G}^{\text{BB}(k)}$}\\
%         {update $\Upsilon^{(j)}=\Upsilon^{*}$ and $j = j+1$}
%         }
%         {update $k:=k+1$}
%     }
%    \caption{Iterative AO algorithm}\label{algo:Algo1}
% \end{algorithm}

Both subproblems \eqref{eq:P6_sub1} and \eqref{eq:P6_sub2} are SOC programs (SOCPs) and can be solved optimally. Their combination yields the AO procedure in Algorithm~1, which alternately updates the baseband precoders and {baseband} combiner until convergence. 

\subsection{Computational Complexity and Convergence Analysis}
Note that \eqref{eq:P6_sub1} and \eqref{eq:P6_sub2} are convex subproblems solved alternately with respect to $\{\mathcal{F}^{\text{BB}}, \mathbf{G}^{\text{BB}}\}$ and $\mathbf{W}_{\text{BB}}$, respectively. Since $F^{(n)}(\cdot)$ is a global under-estimator of $f(x,y)$ that is tight at $(x^{(n)},y^{(n)})$, each subproblem is a convex restriction of \eqref{eq:P6} for which the current point remains feasible; hence, every SCA and AO iteration yields a non-decreasing objective value. The process continues until convergence or the maximum number of iterations, $I_{\rm max}$, is reached. {As the objective function is upper bounded, the sequence of objective values generated by Algorithm~1 converges.} Nevertheless, the obtained solution is generally suboptimal for the original problem \eqref{eq:Orig_P1}.
% Note that the objective function maximization in subproblems \eqref{eq:P6_sub1} and \eqref{eq:P6_sub2} with respect to block of variables $\{\mathcal{F}^{\text{BB}}, \mathbf{G}^{\text{BB}}\}$ and $\mathbf{W}_{\text{BB}}$, respectively, return optimal values for the given initial value. In every iteration $k$, the algorithm returns the improved value of the objective function. According to the algorithm, \eqref{eq:P6_sub1} and \eqref{eq:P6_sub2} are solved iteratively until the convergence of the objective function value or the maximum number of iterations, $I_{\rm max}$, is reached,
% whichever first. Consequently, Algorithm~1 generates a monotonically increasing sequence of the objective values upper bounded by the maximum rate achieved by the considered system without interference. The sequence saturates after a few iterations, and thus, the algorithm converges. However, note that the solution obtained via Algorithm~1 is sub-optimal for the original problem \eqref{eq:Orig_P1}.

Note that both \eqref{eq:P6_sub1} and \eqref{eq:P6_sub2} are SOCPs, so we use the result in \cite{loboLAA1998}. {Since the RF beamformers are fixed, the effective channels are computed once with complexity $\mathcal{O}(B^2M_bM_{\text{h}})$, after which both subproblems depend only on $B$: each has $\mathcal{O}(B^2)$ real variables and $\mathcal{O}(B)$ SOC constraints of dimension $\mathcal{O}(B)$. Hence, the worst-case complexity per SCA iteration is $\mathcal{O}\big(B^{6.5}\log(1/\epsilon)\big)$, where $\epsilon$ is the solution accuracy.}

\begin{figure*}[t]
\centering
\includegraphics[width=\textwidth]{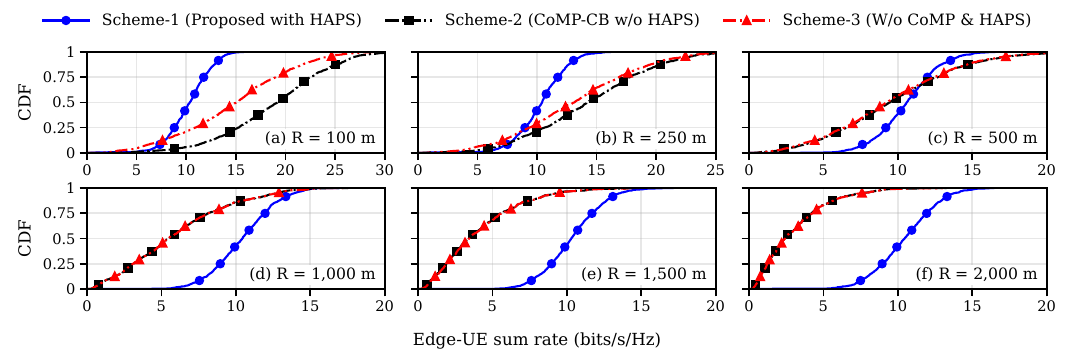}
\caption{{CDFs of the edge-UE sum rate (over feasible realizations) for varying cell radii $R\in\{100, 250, 500, 1000, 1500, 2000\}$ m.}}
\label{fig:CDF_vs_Radius}
\vspace{-0.15in}
\end{figure*}

\begin{table}[t]
\renewcommand{\arraystretch}{1.03}
\caption{Simulation Parameters}
\label{tab:system_parameter}
\centering
\footnotesize
\begin{tabular}{c||c}
\hline
\bfseries Parameters & \bfseries Value\\
\hline
\hline
No. of antennas at BS  & $M_{b_x}$ = 8, $M_{b_z}$ = 8 \\
\hline
No. of antennas at HAPS  & $M_{\text{h}_x}$ = 10, $M_{\text{h}_y}$ = 10 \\
\hline
No. of NLoS paths  & $L$ = 5 \\
\hline
Maximum transmit power & $P_{\text{max},b}$: $45$ dBm, $P_{\text{max},\text{h}}$: {$50$ dBm}\\
\hline
HAPS height & 22 km \\
\hline
{Carrier frequency, bandwidth} & $28$ GHz, {$5$ MHz}\\
\hline
3 dB angles for HAPS{~\cite{3GPP_TR38901}} & {$\varphi^x_{\text{3dB}}=65^\circ$, $\varphi^y_{\text{3dB}}=65^\circ$} \\
\hline
3 dB angles for BS{~\cite{3GPP_TR38901}} & {$\varphi_{\text{3dB}}=65^\circ$, $\theta_{\text{3dB}}=65^\circ$} \\
\hline
Noise temp.{, power} & $T{=}300$ K{, $k_{\text{B}}T{\cdot}5$\,MHz}\\
\hline
{Max. element gain, side-lobe level} & {$8$ dBi, $A_m = 30$ dB ~\cite{3GPP_TR38901}}\\
\hline
{HAPS links (dense urban)~\cite{3GPP_TR38811}} & \begin{tabular}[c]{@{}c@{}}{Rician $K=10$ dB, free-space loss,}\\ {$\sigma_{\text{SF}}=0.6$--$2.9$ dB}\end{tabular}\\
\hline
{BS--UE path loss (dB)~\cite{3GPP_TR38901},} & {NLoS: $32.4+20\log_{10}f_c^{\text{GHz}}$}\\
{$D$: distance (m)} & {$+31.9\log_{10}D$, $\sigma_{\text{SF}}=8.2$ dB}\\\hline
\end{tabular}
\end{table} 

\section{Numerical Results}
{In this section, we evaluate the proposed Algorithm~1 via numerical simulations using two metrics: the {cumulative} distribution functions (CDFs) of the edge-UE sum rate and the outage probability, i.e., the fraction of infeasible realizations.} In all simulations, {we consider $B=2$ terrestrial cells} of radius $R$, each with a central BS, one near UE, and one edge UE, located in a 30-km-radius urban area {(within the 200-km HAPS footprint)} with many skyscrapers causing heavy shadowing near the cell edge, with the FD-HAPS above the area center. {Each result uses $50$ topologies $\times$ $20$ i.i.d. channel realizations, with}  $r_{\text{QoS}}=0.1$ bits/s/Hz. Table~\ref{tab:system_parameter} lists the parameters and their values used in the simulations~\cite{Lin-Lin-Huang-Cola-Zhu-TSP-2019}.

We refer to the proposed approach as Scheme-1, ``Joint design with HAPS (with NOMA)'', and compare it with two benchmarks: (i) Scheme-2, ``Joint design without HAPS (with NOMA)'', i.e., terrestrial {coordinated multipoint with coordinated beamforming (CoMP-CB) without HAPS \cite{gesbert2010IEEEjsac}, where the BSs {share CSI but not user data} to manage interference at the cell-edge UEs and serve their near and edge UEs via NOMA}; and (ii) Scheme-3, ``Uncoordinated design without HAPS (with NOMA)'', which is identical to Scheme-2 except that the BSs do not coordinate.
% \begin{figure}[h]
% \centerline{\includegraphics[width=0.6\linewidth]{convergence_10x10_R200m.png}}
% \caption{Conversion behaviour of the proposed SCA-based AO  Algorithm~1.}\label{fig:convergence_10x10_R200m}
% \vspace{-0.1in}
% \end{figure}
% We first discuss the convergence behaviour of the proposed SCA-based AO algorithm. Fig.~\ref{fig:convergence_10x10_R200m} plots the objective function values achieved by the Algorithm~1 after each SCA iteration for three different random channels and transmit power $P_{\text{max}}$ values. It can be observed that the algorithm converges within ten iterations for all three channels.

% \begin{figure}[h]
% \centerline{\includegraphics[width=1\linewidth]{Typical_locations.png}}
% \caption{The top view of an example locations of the FD-HAPS, terrestrial BSs with thier near and edge UEs.}\label{fig:topology_example}
% \vspace{-0.1in}
% \end{figure}
Fig.~\ref{fig:CDF_vs_Radius} compares the CDFs of the edge-UE sum rate of the three schemes over the feasible channel realizations for $R$ from 100 m to {$2{,}000$ m}. First, the edge-UE sum rate of Scheme-1 remains nearly unchanged with $R$ because the LoS BS-to-HAPS and HAPS-to-UE links are almost independent of the cell size. Moreover, it does not increase with $R$. This is because the closely spaced edge UEs share one HAPS beam footprint, so their mutual interference persists for all $R$. Second, the edge-UE sum rates of Scheme-2 and Scheme-3 decrease sharply with $R$ due to the higher path loss and shadowing of the terrestrial links, and nearly coincide for {$R \geq 1{,}000$ m} as the ICI vanishes; hence, Scheme-1 outperforms both for $R \geq 500$ m, by about {3.5 times at $R=2{,}000$ m}. Third, at {$R\leq 250$ m}, Scheme-2 and Scheme-3 achieve higher edge-UE sum rates, but at the expense of much lower near-UE rates.

\begin{figure}[h]
\centerline{\includegraphics[width=1\linewidth]{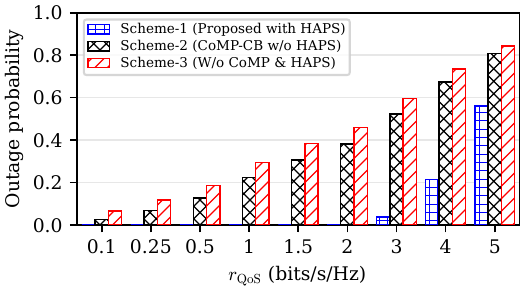}}
\caption{{Outage probability versus $r_{\text{QoS}}$ for $R = 500$ m.}}\label{fig:outage_vs_rqos}
\vspace{-0.15in}
\end{figure}
Fig.~\ref{fig:outage_vs_rqos} shows the outage probability versus $r_{\text{QoS}}$ for $R=500$ m. Scheme-1 {is essentially outage-free} up to $r_{\text{QoS}}=2$ bits/s/Hz, whereas Scheme-2 and Scheme-3 already reach about {38\% and 46\%}, respectively, because the FD-HAPS gives each edge UE an additional LoS link that bypasses terrestrial shadowing. {Scheme-2} outperforms Scheme-3 due to BS coordination.

% \begin{figure}[h]
% \centerline{\includegraphics[width=0.8\linewidth]{sumrate_vs_power_10x10.png}}
% \caption{Sum rates versus transmit power of the BS with varying terrestrial cell radii $\in \{100, 500, 1000 \}$ m.}\label{fig:sumrate_vs_power_10x10}
% \vspace{-0.1in}
% \end{figure}
% Fig.~\ref{fig:sumrate_vs_power_10x10} shows the effect of BS transmit power on the system and edge-UE sum rates of Scheme-1 for different terrestrial cell radii. As the maximum BS transmit power increases, the edge-UE sum rate first rises and then saturates due to mutual interference between the edge UEs. For near UEs, a larger cell radius reduces HAPS-beam leakage and, hence, interference. Therefore, the system sum rate increases with both cell radius and transmit power.
\section{Conclusion}
We studied hybrid beamforming for a HAPS-assisted multi-cell network, maximizing the edge-UE sum rate under power and QoS constraints via SOCP-based approximations. Simulations showed that{, unlike terrestrial-only baselines, the proposed HAPS-complemented scheme sustains the edge-UE sum rate and avoids outage as the cell radius grows}. Future work will consider residual SI, imperfect CSI, larger networks, and half-duplex and orthogonal multiple access baselines.

\bibliography{ref_haps.bib}
\bibliographystyle{IEEEtran}
\end{document}